\documentclass[letterpaper,twocolumn,10pt]{article}
\usepackage{usenix,epsfig}
\graphicspath{{./images/}}
\usepackage{listings}
\usepackage{url}
\usepackage{amsmath}
\usepackage{graphicx}
\usepackage{booktabs}
\usepackage{epstopdf}
\usepackage{caption}
\usepackage[finalnew]{trackchanges}

\addeditor{Yan} 
\lstdefinestyle{myPrologstyle}
{
    language=Prolog,
    breaklines=true,
    basicstyle = \ttfamily\color{blue},
    moredelim = [s][\color{black}]{(}{)},
    literate =
        {:-}{{\textcolor{black}{:-}}}2
        {,}{{\textcolor{black}{,}}}1
        {.}{{\textcolor{black}{.}}}1
}

\begin{document}
%
%don't want date printed
%\date{\today}

%for single author (just remove % characters)
 \author{
 {\rm Yan Shvartzshnaider}\\
 NYU
 \and
 {\rm Schrasing Tong}\\
 Princeton University
 \and
 {\rm Thomas Wies}\\
 NYU
 \and
 {\rm Paula Kift}\\
 NYU
 \and 
 {\rm Helen Nissenbaum}\\
 NYU
 \and
 {\rm Lakshminarayanan Subramanian}\\
 NYU
 \and
 {\rm Prateek Mittal}\\
 Princeton University
 } % end author

  \title{\Large \bf Crowdsourced, Actionable and Verifiable Contextual Informational Norms}
  %\runningtitle{Crowdsourcing Verifiable Contextual Integrity Norms}

  %\subtitle{...}
\maketitle

\subsection*{Abstract}

There is often a fundamental mismatch between programmable privacy frameworks, on the one hand, and the ever shifting privacy expectations of computer system users, on the other hand. Based on the theory of contextual integrity (CI)~\cite{nissenbaum2015respecting}, our paper addresses this problem by proposing a privacy framework that translates users' privacy expectations (norms) into a set of actionable privacy rules that are rooted in the language of CI. These norms are then encoded using Datalog logic specification to develop an information system that is able to verify whether information flows are appropriate and the privacy of users thus preserved. A particular benefit of our framework is that it can automatically adapt as users' privacy expectations evolve over time. 

To evaluate our proposed framework, we conducted an extensive survey involving more than 450 participants and 1400 questions to derive a set of privacy norms in the educational context. Based on the crowdsourced responses, we demonstrate that our framework can derive a compact Datalog encoding of the privacy norms  which can in principle be directly used for enforcing privacy of information flows within this context. In addition, our framework can automatically detect logical inconsistencies between individual users' privacy expectations and the derived privacy logic.

%!TEX root = main.tex
\section{Introduction}

Incorporating privacy expectations into real world systems remains an important research challenge. For the privacy-by-design initiative~\cite{privacy-by-design-workshop} this calls for more than technical rigor; it calls for the adoption of a socially meaningful conception of privacy, namely, one that meets people's expectations, and is ethically and legally legitimate. In the case of online information platforms, where privacy must be maintained amidst complicated flows among participants, and between the participants and the platform, the challenge is particularly acute~\cite{reidenberg2014disagreeable,madejski2012study,breaux2009distributed}. In this regard, the account of privacy~\cite{nissenbaum2011contextual} as contextual integrity has been promising, inspiring work on formal expression of contextual informational norms, on detection of infractions, and on approaches to accountability and enforcement~\cite{chowdhury2013privacy, barth2006privacy,criado2015implicit}. These, and other similar efforts~\cite{gordon2011comparing} have made extensive contributions to the technical field --- generating machine-readable access rules and implementing complex constraints that map given rules. \textit{Building on this body of work, our project incorporates an additional component, namely, the discovery, articulation and verification of norms.}

\begin{figure}
\center
      \includegraphics[width= 0.5\textwidth]{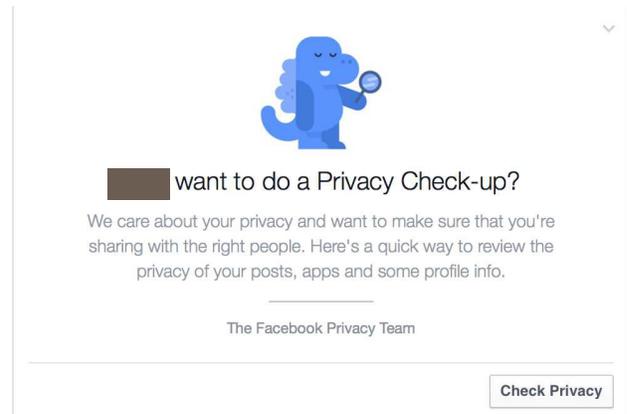}
    \caption{A screenshot from Facebook asking users to check their privacy policies.} 
    \label{fig:facebook}
\end{figure}

The framework we are developing, based on the theory of contextual integrity, offers designers a comprehensive form of this functionality centered on contextual informational norms. This framework not only provides tools to implement privacy rules based on social norms, but also to continuously update those rules according to the fluid and ongoing evolution of privacy expectations, as is characteristic of social norms more generally. In cases where the discovery of norms may not be possible, and in systems where privacy rules are based on legal and policy documents as well as professional codes, our framework meets the remaining challenge faced by designers and requirements engineers alike, namely expressing privacy norms and implementing them as privacy rules on the one hand, and ensuring their consistency and overall compliance with privacy expectations, on the other hand. 

The theory of contextual integrity informs the structure of our privacy rules, and our project embeds these in a logic framework which encompasses a system for learning norms as well as the ability to  enforce them in a system that is built on these norms.
Further, these norms are formally expressed and verified to safeguard from  privacy violations that may result from a semantic gap between privacy norms and mechanisms that enforce the privacy norms. The need for such functionalities is reflected in Figure~\ref{fig:facebook}, depicting Facebook's privacy check-up utility. Our system offers the ability to automate such check-ups. The framework continuously functions in the background and can track the evolution of norms. 

In summary, our privacy framework makes the following key contributions:

\begin{enumerate}

\item \textbf{Formal expression based on the theory of privacy as CI.}  
Our framework uses the theory to contextual integrity to formalize informational norms as logical rules using three key parameters: a) actors (senders, subjects and receivers of information, usually people or organizations); b) attributes (information types); and c) transmission principles. (See Section~\ref{sec:ci} for further explanation.) 

  \item \textbf{A methodology for discovering informational (privacy) norms through crowdsourcing} \add[Yan]{We have developed an approach to identifying contextual privacy norms based on the "wisdom of the crowds," in this case, the collective input from system users/participants.} \remove[Yan]{based on the language of CI.}  \remove[Yan]{We elicit informational norms based on a crowdsourcing approach that asks a large pool of respondents on AMT whether a set of automatically generated privacy statements that are based on the language of CI meets their privacy expectations.}

\item \textbf{Converting crowd-sourced responses to a corresponding
    privacy logic.} To derive a functional privacy logic, our system
  encodes structured informational norms discovered through the  crowdsourcing methodology using the Datalog declarative language.

\item \textbf{Verification of privacy norms.} Our framework is designed to support  formal verification of the derived privacy logic. Specifically, we were able to verify for consistency of flows described by the logic including checking the consistency of transitive flows. 

\end{enumerate}

The paper is organized as follows: Next section provides a brief overview of the CI theoretical framework. Section~\ref{sec:design} describes our framework design.  Section~\ref{sec:vap} discusses how we represent and evaluate CI norms in Datalog, while verifying additional high-level properties using the theorem prover Z3~\cite{de2008z3}. We also describe our crowdsourcing methods in this section. We provide details on the evaluation in Section ~\ref{sec:evaluation} and reflect on our results in Section~\ref{sec:discussion}.   
Finally, in Sections~\ref{sec:related_work} and~\ref{sec:conclusion} we describe related works and then conclude the paper. 
%!TEX root = main.tex
\section{Contextual Integrity Primer }\label{sec:ci}

The theory of contextual integrity (CI) postulates that informational privacy is not all about secrecy (blocking information)~{\cite{posner1977right}} or control~\cite{westin1967privacy} but about the appropriateness of information flow within a particular context. Appropriateness of flow means flow that is compliant with contextual norms governing informational flows. To express an informational norm one must specify key parameters: \emph{actors} (senders, recipients and subjects), \emph{attributes} (the type of information at hand) and \emph{transmission principles} (the constraints imposed on a particular information flow). Taken together, these components constitute context-relative informational norms. For instance, in the health context, the patient, acting in his capacity as both the sender and subject of an information flow, could be telling his doctor, the recipient, about his health issues, the attribute. The information flow would be constrained by the transmission principle of confidentiality, which restricts the onward flow of this information to other parties. Exchange the doctor for a friend, and the transmission principle might be reciprocity instead, since friends tend to expect to hear about each other's problems. A patient, by contrast, does not expect to hear about the health issues of his doctor. This is because the context of health and the context of friendship have different overarching goals: the doctor is there to promote the patient's health; friends are there to support each other. 
An informational norm is breached when an action or practice disrupts the actors, attributes, or transmission principles within a given information flow.  Contextual integrity ``is preserved when informational norms are respected and violated when informational norms are breached''~{\cite{nissenbaum2009privacy}. In legal deliberations about privacy, the same intuition is captured by the concept of ``reasonable expectation of privacy''~\cite{solove2003information}.

It is worth emphasizing one of the most fundamental aspects of CI, namely that, in order to determine whether or not information flows respect or violate privacy expectations within a given context, one must address all three parameters: actors, information types, and transmission principles. Omitting any one of them may lead to an inconclusive or ambiguous description. Accordingly, any formal rendering of information flows, for the purpose of assessing their appropriateness, needs to include independent variables for these parameters. 

CI recognizes that informational norms, like other social norms, are constantly evolving. Sometimes, changes in a sociotechnical environment, such as the ones we are experiencing in this ``digital age", can be quite rapid. Although the theory of CI has a prima facie preference for entrenched informational norms, it also allows for normative transformations when the resultant norms can better promote the values, goals, and ends of a given context.  People learn and adopt implicit and explicit informational norms from their families, friends, and communities; by watching how people behave and how they react to other people's behavior; from educational training, the arts, and cultural activities; from the study of law and policy, and so forth. 

Designing a system that translates contextual informational norms into privacy rules must either depend on a range of legitimate external sources of knowledge (e.g. social scientists, ethicists, law, professional codes, etc.), or must incorporate, internally, some form of norm discovery functionality.
\medskip\\
\textit{Our project has adopted CI as our underlying conception of privacy. There is sufficient regard for it in the privacy community to consider this reasonably uncontroversial. However, readers who are interested in learning about its underlying rationale as well as its policy applications, might wish to consult the relevant literature~\cite{nissenbaum2009privacy, website:white_house_report}. In Section~{\ref{sec:related_work}}, we describe related research that pursues similar ends utilizing alternative conceptions.}

%!TEX root = main.tex
\begin{figure}
\center
      \includegraphics[width= 0.5\textwidth]{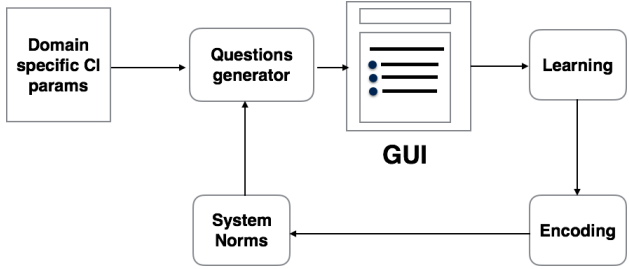}
    \caption{Our framework is designed to support several system states. The overall operation consists of 1) generating questions that correspond to information flows, 2) asking the questions, 3) deriving insights from the answers, 4) deciding on which set of actionable privacy rules (APR) to enforce, 5) generating questions from the enforced norms, 6) repeating by returning to step 2. 
 } 
    \label{fig:components}
\end{figure}

\section{\change[Yan]{Framework Design}{CI-based Privacy Framework}}\label{sec:design}

The key components of our CI-based privacy framework is described in
Figure~\ref{fig:components}. 
Our framework  is adaptable to a range of information systems and platforms. We have conceived it as an independent service that can be integrated into the functionality of these respective platforms and  systems.
Our CI-based framework takes as input a simple state space description of a domain
based on the contextual integrity definitions. Specifically, this implies, 
the input includes the list of actors, subjects, information types and transmission principles that are relevant to the domain of interest. In addition, the domain
specific input parameters may also provide information about the importance of 
specific attributes and subjects that correspond to common information flows in the domain to guide the privacy logic generation process. Given the domain specific input, our framework performs three key steps:

\begin{description}
\item[Norm discovery through crowdsourcing.] The Question Generation (QG) stage takes as input only the list of parameters associated with a context, including the list of actors, information types, and transmission principles. The QG generates questions from the prevailing set of privacy norms which cover the common information flows. These questions are crowdsourced to a set of users who provide a simple ``Yes/No/Irrelevant" response
for each question. Based on the users' collective responses, the framework will derive the  corresponding norms.
\item[Learning and Encoding the Privacy logic.]At the Learning stage, the answers to the questions are analyzed to learn users' privacy expectations. Here, we are specifically learning a collective notion of privacy expectation within a domain based on a consensus metric defined on the individual user responses to each question. Next, at the Encoding stage, the norms reflected by the answers to each question are encoded into a set of actionable privacy rules using Datalog: they form the logic behind the users' privacy expectation. The logic is  passed on to the System state which is responsible for privacy norm verification and enforcement based on internal configurations, which may or may not take users' perception logic into account.

\item[Verification.] All of the information flows that form part of the system are governed by privacy rules. Using the Datalog encoding, the system evaluates every information flow against the prevailing set of privacy rules. Furthermore, the system is capable of verifying other meta-assertions for the derived privacy logic, including validating that the enforced norms are consistent and disapproved information flows are impermissible.

\end{description}

In addition, the framework is designed to support learning of constantly evolving norms; as it traverses through states, new norms are introduced and old ones are re-evaluated. \remove{Norms that at some point are discarded and not approved can regain ``popularity'' as more users join and use the system. 
In the next section, we describe the process behind each of the components in greater detail. }

\subsection{An example: the educational context}

As a case study, we chose to apply our framework to the educational context, since this is the context with which we have the most familiarity; the discussion, however, can easily be extended to other contexts. We outline a specific example
of actors, information types and transmission principles in the educational context below:

\begin{description}
\item[Actors (Senders, Recipients, Subjects):] Students, Professors, TAs, Registrar, University IT staff, academic advisor

\item[Examples of Attributes:] Grades, Transcript, Name, Email address, Address, Record of attendance, Level of participation in class, Photo, Library records, Contents posted on online learning systems (e.g., Blackboard, Classes, etc.), term paper

\item[Example Transmission principles:]\quad
  \begin{description}
  \item[Knowledge:] If the $\langle$ sender $\rangle$ let the $\langle$ subject $\rangle$ know
  \item[Permission:] If the $\langle$ sender $\rangle$ asked for the  $\langle$ subject's $\rangle$ permission
  \item[Breach of contract:] If the $\langle$ subject $\rangle$ is performing below a certain standard
  \end{description}
  \end{description}
  
We elaborate upon this example in Section~\ref{sec:crowdsrouced} in greater detail to show how to automatically design the question generator given a state space description.

%\input{system}
%!TEX root = main.tex
\section{Verifiable  and Actionable Privacy Rules}\label{sec:vap}
\add[Yan]{In this section, we first describe the encoding and verification of the privacy logic and then proceed to describing the procedure behind the crowdsourcing of norms. }

\subsection{Representation and verified enforcement}\label{sec:datalog}

%======== Copied from the Proposal =====
\label{sec:datalog}

We use the declarative programming language Datalog~\cite{ceri1989you}
to formally represent contexts, for evaluating the privacy rules in
these contexts, and for automating critical aspects of the crowdsourced
learning component.
%
%\paragraph{Rationale for using Datalog.}
Datalog is a well-studied language and formalism that has found
numerous applications, including the analysis of social
networks~\cite{DBLP:journals/tkde/SeoGL15} and as a language for
expressing privacy and security
policies (see, e.g.,
~\cite{DBLP:conf/sp/DeTreville02,DBLP:conf/padl/LiM03,DBLP:conf/datalog/Bonatti10}).  
%Its relevance for the formalization of contextual integrity has been noticed before~\cite{DBLP:conf/trustbus/LamMS09}. 

Our main motivation for building our formalization on Datalog can be summarized
as follows:
\begin{itemize}
\item The encoding of contextual privacy rules to Datalog is elegant
  and easily understandable. In particular, we can leverage Datalog's
  query mechanism to automatically evaluate and check the properties of privacy norms in concrete contexts. We provide several
  examples below.
\item Datalog forms a fragment of first-order predicate logic.  Hence,
  the language's semantics are well understood. In particular, this
  allows us to use powerful first-order theorem provers to
  automatically prove high-level privacy properties of the norms that
  we learn for a specific context.
\item Datalog provides a good trade-off between expressiveness and
  complexity. In particular, Datalog's computational model is not
  Turing complete, which means that all queries are guaranteed to
  terminate. We mostly restrict ourselves to a specific fragment of
  Datalog referred to as \emph{unions of conjunctive queries}. This
  fragment is well-studied in the AI, logic programming, and database
  communities. In particular, efficient algorithms for its treatment
  have already been developed~\cite{DBLP:journals/tcs/FaginKMP05}.
\end{itemize}

\subsection{Encoding Contextual Privacy in Datalog}\label{sec:encoding}

Datalog is a fragment of the logical programming language Prolog.  A
Datalog program consists of \emph{clauses} that define predicates on
entities. Our Datalog encoding of contextual privacy specifies predicates on
entities that stand for contexts, actors, attributes, and transmission
principles as described in Section~\ref{sec:design}. Central to the
encoding is the predicate.
\begin{verbatim}
allowed(Ctx, Sndr, Recp, Subj, Attr, Tr)
\end{verbatim}
This predicate models that in context \texttt{Ctx}, actor
\texttt{Sndr} is allowed to send information on attribute
\texttt{Attr} of actor \texttt{Subj} to actor \texttt{Recp} under
transmission principle \texttt{Tr}.  For example, the following fact
states that in the classroom context (denoted by \texttt{class}),
\texttt{bob} can send his own grade to \texttt{alice} with
transmission principle \texttt{confidentiality}.

\begin{verbatim}
allowed(class, bob, alice, bob, grade, 
        confidentiality).
\end{verbatim}

In order to be able to express the privacy rules, we introduce a
ternary predicate \texttt{inrole(Context, Actor, Role)}, which models
that in the given context, the given actor is in the specified
role. For example, the following fact states that in the classroom
context, \texttt{bob} is in the role of \texttt{student}
\begin{verbatim}
inrole(class, bob, student).
\end{verbatim}

The predicate \texttt{allowed} captures the rules of all privacy
contexts. The individual rules for each context are stated using
clauses that contain the predicate \texttt{allowed} in the head. As an
example, the following clause codifies the rule that professors can
let any student know her own grade upon her request:
\begin{verbatim}
allowed(class, Sndr, Recp, Subj, grade, 
        need) :-
  inrole(class, Sndr, professor),
  inrole(class, Recp, student),
  Subj = Recp.
\end{verbatim}
Note the close correspondence between this clause and the shape of the
survey questions discussed in Section~\ref{sec:design}.

A Datalog program is executed by evaluating a \emph{query} that asks
whether a certain conjunction of predicates holds true according to
the clauses in the program. Suppose that in our social platform we
have a classroom context whose actors are described by the following
facts:
\begin{verbatim}
inrole(class, bob, student).
inrole(class, alice, student).
inrole(class, steve, professor).
\end{verbatim}
Then we can use the query mechanism to check whether a specific
information flow satisfies all the specified privacy rules. For
instance, given the above privacy rule and facts, the query
\begin{verbatim}
?- allowed(class, steve, bob, bob, grade, 
           need).
\end{verbatim}
evaluates to \texttt{true}, indicating that the corresponding
information flow is admissible. On the other hand, the query
\begin{verbatim}
?- allowed(class, steve, alice, bob, 
           grade, need).
\end{verbatim}
evaluates to \texttt{false}, indicating that this flow is not
permitted.
%
\iffalse
Queries can also contain variables. These variables are
implicitly existentially quantified. We can use this feature to
check automatically whether the privacy rules satisfy desirable
privacy properties. Consider, e.g., the following query
\begin{verbatim}
?- allowed(class, X, Y, Z, grade, confidentiality), 
   allowed(class, Y, W, Z, grade, T).
\end{verbatim}
This query asks whether it is possible, under the specified rules,
for an actor \texttt{Y} to forward \texttt{Z}'s grade after
having received it confidentially. If the query returns
\texttt{false}, then this transitive flow of information is disallowed.
\fi

\subsection{Privacy logic verification}\label{lbl:verification}

Datalog allows us to formally specify CI rules in a declarative
manner. In particular, we can use the query mechanism of a Datalog
interpreter to check that all information flows in a social platform
are consistent with the specified rules. The semantics of Datalog
guarantees that no norm-violating flows will be permitted at
\emph{run-time}. The fact that Datalog provides a formally defined
semantics for CI rules has another important advantage: it enables us
to verify that the rules of a given context satisfy desirable
high-level privacy properties that are not immediately evident from
the rules. For example, we may want to verify that our rules do not
permit the flow of confidential information to third parties. We can
check such properties \emph{statically} before the rules are in
effect. 

One specific application of formal verification in our proposed
crowd-sourced learning approach is that we can check whether the rules
that have been approved by the crowd are not violating the rules that
have been disapproved. In particular, we can leverage formal
verification to facilitate the automatic adjustment of the threshold
that determines which rules are considered to be approved.
In the following, we describe in more detail how such static
verification tasks can be automated.

The problem of verifying that a given set of rules $\mathcal{R}$
satisfies a given property $P$ amounts to checking logical validity of
the implication $\mathcal{R} \Rightarrow P$, or dually, that the
conjunction $\mathcal{R} \land \neg P$ is unsatisfiable. For simple
properties $P$, the latter can be checked directly using Datalog
queries. However, in general, Datalog queries are not sufficiently
expressive to verify complex high-level properties. Fortunately, we
can embed Datalog in a more expressive logic that is still
amenable to automated reasoning and yields tractable performance for
the static verification of high-level properties in
practice.

Datalog is a fragment of first-order predicate logic. Specifically,
suppose we are given a set of rules $\mathcal{R} = \{R_1,\dots,R_n\}$
where each rule $R_i$ is specified by a Datalog clause of the form
\[
\begin{array}{l}
\mathtt{allowed}(\mathit{C}, \mathit{Sn}, \mathit{R}, \mathit{Su}, \mathit{A}, \mathit{T}) \mathrel{\texttt{:-}}
C_{i,1}, \dots, C_{i,m_i}.
\end{array}
\]
and the atoms $C_{i,j}$ are either \texttt{in\_role} predicates over
the given variables in the head of the clause or equalities between
these variables and constants such as \texttt{student},
\texttt{grade}, etc. Then the semantics of these clauses is captured
by the following quantified formula:
\[
\begin{array}{l}
  \forall \mathit{C}, \mathit{Sn}, \mathit{R}, \mathit{Su},
  \mathit{A}, \mathit{T}.\\
  \quad \mathtt{allowed}(\mathit{C}, \mathit{Sn}, \mathit{R},
  \mathit{Su}, \mathit{A}, \mathit{T}) \Leftrightarrow\\
  \quad \quad (C_{1,1} \land \dots \land C_{1,m_1})
    \lor \dots \lor (C_{n,1} \land \dots \land C_{n,m_n})
\end{array}
\]
This formula falls into \emph{Effective Propositional Logic} (EPR), a
decidable fragment of first-order predicate
logic~\cite{borger2001classical}. Several automated theorem provers
implement decision procedures for EPR, e.g., the Satisfiability Modulo
Theories solver Z3~\cite{de2008z3}. If the privacy property $P$ of
interest is itself expressible in EPR, then we can use Z3 to
automatically check that the rules $\mathcal{R}$ guarantee
$P$. Fortunately, many properties of interest are indeed expressible
in EPR. For example, the following EPR formula expresses that in the
classroom context, a professor should not be allowed to send a
student's grade to any other student, unless that other student is a
TA:
\[ 
\begin{array}{l}
  \forall \mathit{Sn}, \mathit{R}, \mathit{Su}, \mathit{T}.\\
  \quad \mathtt{in\_role}(Sn,\mathtt{professor}) \land \mathtt{in\_role}(R,
    \mathtt{student}) \land {}\\
  \quad \mathtt{allowed}(\mathit{Sn}, R, \mathit{Su}, \mathtt{grade},
  T) \Rightarrow \\
  \quad \quad \mathit{Su} = R \lor \mathtt{in\_role}(R, \mathtt{TA})
\end{array}
\]
Note that this property cannot be checked with a simple Datalog
query. We need the additional expressiveness provided by EPR.  More
generally, EPR can express properties about transitive information
flows that involve arbitrarily long sequences of information
exchanges. Since the satisfiability problem for EPR is decidable, we
can verify these properties fully automatically using tools such as
Z3. In particular, if a specified property is not guaranteed by the
rules, the theorem prover will produce a model describing an
information flow that respects the rules but violates the
property. Using this model, we can then identify the rules that are
responsible for this property violation.

\subsection{Crowdsourcing privacy rules}\label{sec:crowdsrouced}
In this section we describe the procedure for constructing questions from CI norms. 
Users are provided with a series of multiple-choice questions; each possible answer corresponds to a different information flow. This allows us to establish a baseline for users' expected privacy preferences. To simulate this process, we relied on Amazon Mechanical Turk (AMT), as we describe in Section~\ref{sec:evaluation}.

\subsubsection{Constructing questions}\label{sec:constructNoms}

\begin{figure}
\centering
      \includegraphics[width=0.5\textwidth]{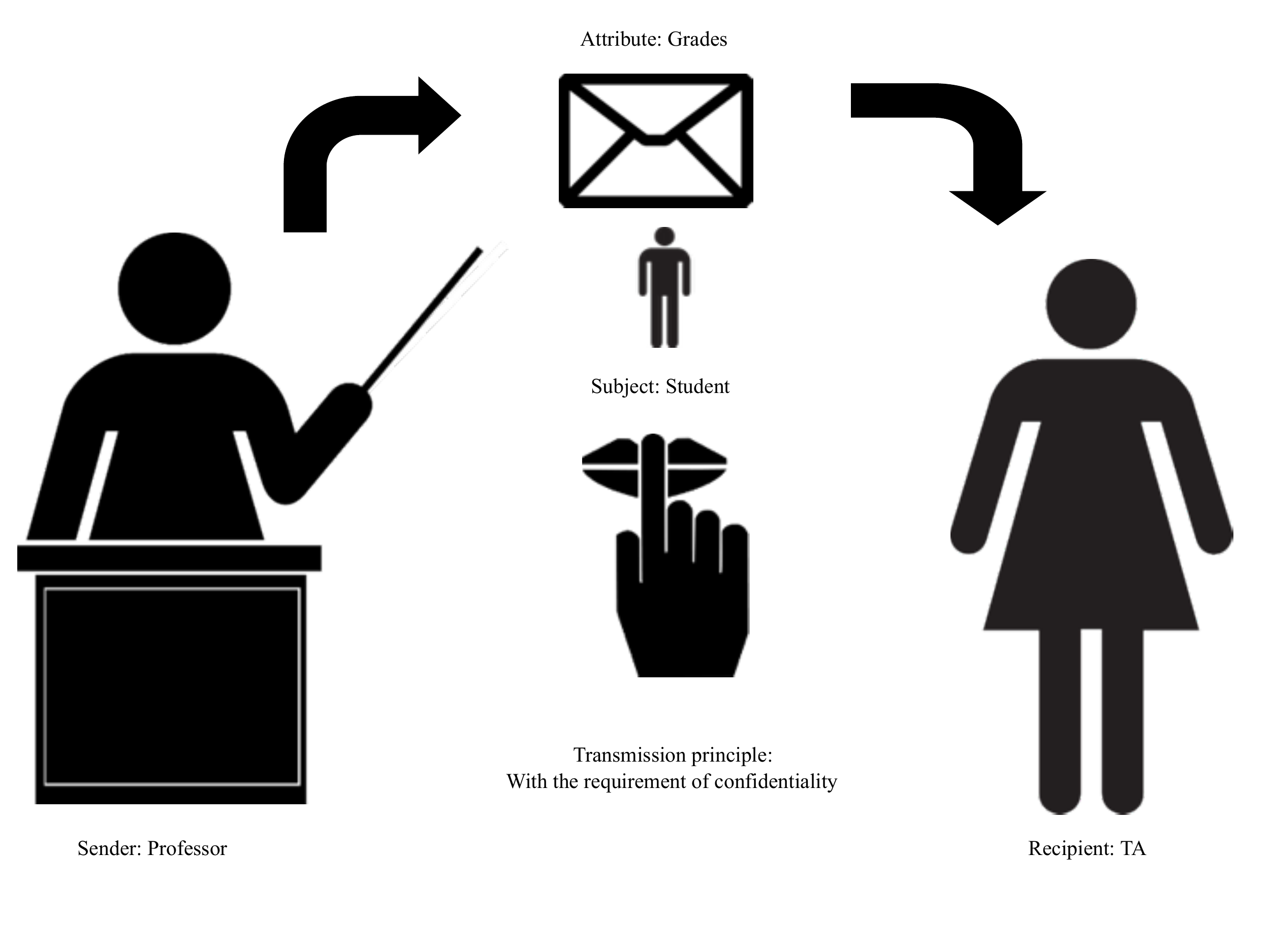}
    \caption{Information norm: The professors is allowed to share the student's grade with the student's TA with the requirement of confidentiality.}
    \label{fig:CINorm}
\end{figure} 

As depicted by the example in Figure~\ref{fig:CINorm}, CI norms contain the following elements: \emph{actors} (senders, subjects and recipients), \emph{attributes} (information types), and \emph{transmission principles} (constraints on flow).
Below are some examples of their possible values. We should note that this is not a comprehensive set of values. Rather, we use this preliminary set to bootstrap the system. We discuss how new attribute values can be added in the next Section.

\begin{description}
\item[Senders:] Professors, TAs, Registrar, University librarians, University IT staff, classmates, academic advisor

\item[Subjects:] Student

\item[Recipients:] Professors, TAs, Registrar, University librarians, University IT staff, Department chair, Classmates, Parents, Academic advisor

\item[Attributes:] Grades, Transcript, Name, Email address, Address, Record of attendance, Level of participation in class, Photo, Library records, Contents posted on online learning systems (e.g., Blackboard, Classes, etc.), term paper

\item[Transmission principles:]\quad
  \begin{description}
  \item[Confidentiality:] With the requirement of confidentiality
  \item[Knowledge:] If the $\langle$ sender $\rangle$ let the student know
  \item[Permission:] If the $\langle$ sender $\rangle$ asked for the student's permission
  \item[Purpose:] To improve the learning experience
  \item[Breach of contract:] If the student is performing below a B- standard
  \item[Need:] If requested by the $\langle$ recipient $\rangle$
  \end{description}
\end{description}

\add[Yan] {Note that, as a theoretical framework, CI does not mandate any particular implementation. Although the resulting 5-tuple is a conventional way of representing different information flows, it can be extended to include additional elements (e.g., by introducing the \emph{context} element in our encoding) to fully encompass the  expressive reality. For the sake of simplicity we rely on the 5-tuple format for now.} 
We inject each of the elements into the following Yes-or-No question template: \\

{\small ``\emph{Is it acceptable for the $\langle$sender$\rangle$ to share the $\langle$subject$\rangle$'s $\langle$attribute$\rangle$ with $\langle$recipient$\rangle$$\langle$transmission principle$\rangle$?"}} \\

From the answers to the crowdsourced questions we extracted a truth table describing the information flows that are admissible in this context. 
The admissibility is determined by a rank-threshold which is set in the system configurations.  The learned truth table represents a formal specification of the contextual privacy norms.

Note that during question generation we do not enumerate the full  space of all possible privacy norms that can be expressed over the given parameter values. Instead, we rely on the input of a privacy expert to reduce the explored space to those candidate norms that cover the bulk of the relevant information flows. For the educational context, expert knowledge enabled us to reduce the relevant space from an initial 28 thousand questions to only 1411.

\subsubsection{Introduction of new norms}\label{sec:evolution}

The \change{OPR}{crowdsourced} set of privacy norms will vary depending on the initial input; however, for the system to evolve (i.e., in order for it to be able to introduce new privacy rules) it needs to be able to adapt to information flows that have not previously been entered into the system.  We envision the system to evolve in the following manner:

\begin{description}
\item[User input.] In order to adapt to new information flows, and to be able to develop rules that correspond to them, users will in some cases be able to ``edit" their answer to a particular question, or even the question itself. For example, the \emph{transmission principle} can be used to indicate a new actor, e.g.,  ``\emph{With the permission of the [NEW ACTOR]}", so a question can be displayed as follows: \emph{``Is it acceptable for the student's professor to share the student's grades with the student's TA  \textbf{with permission of the [NEW ACTOR]}?"} Users will be able to either choose from a selection of preexisting actors or to introduce a new actor to the system. Answer options that correspond to ``Does not make sense'' will weed out spam lows. Other CI attributes, such as Subject, Sender, Receiver, etc., can be introduced in a similar fashion. 

\item[Expert input.] As an alternative or supplementary approach  to the crowd-sourced one, new actors could also be introduced by designated domain experts. Experts periodically decide, based on external factors, which new actors/attributes should be introduced in to the system. 

\end{description}

Both approaches can be used in combination, in a curation-type mode. The expert could check the inputs from users to make sure they are relevant to a given context. 

Ultimately, privacy rules corresponding to new elements will trigger the generation of new corresponding questions which will be presented to the users for ranking. \remove{The ones that get ranked over the decided threshold will be integrated into the set of norms that is enforced by the system. }

%!TEX root = main.tex
\section{Evaluation}\label{sec:evaluation}

In our experiments we aim to evaluate the following components:
\begin{itemize}
\item How the metrics we propose can serve as indicators of the state of norms that have already been approved and whether users are satisfied with the socially derived Actionable Privacy Rule (APR) set 
\item Test our automatic verification approach for consistency of the derived privacy logic 
\end{itemize}

\subsection{\change[Yan]{Survey}{Simulation} design}
\remove[Yan]{We begin by describing the survey and the metrics used to select
operational norms based on the feedback provided to the survey
questions.}

For the purpose of this simulation, we took the educational context as an example to test whether the CI framework would be able to better encapsulate users' privacy expectations.  We constructed a context-specific set of \change[Yan]{survey}{questions} that would allow us to crowdsource corresponding informational norms. Our target population was US residents, between 18-26 years of age, and currently enrolled in (or graduated within the past three years from) an institution of higher education in the United  States.  \change[Yan]{We conducted these surveys using an online survey designed with Qualtrics and administered on Amazon's Mechanical Turk.  }{We posed these questions using an online survey designed with Qualtrics and administered on Amazon's Mechanical Turk. }

We used a \remove[Yan]{python} script \remove[Yan]{(see Appendix)}  to generate the initial set of
norms based on the most common CI parameters in a classroom setting
(e.g., teachers and students as actors; grades as attribute; knowledge
or consent as transmission principles).  To decrease the total number
of questions asked, \change[Yan]{an expert}{two of the authors} performed a preliminary scan of the
norms to identify the ones that clearly did not make any sense. Rather
than manually going through the questions one by one, \change[Yan]{the expert}{the authors} 
focused exclusively on valid pairs of senders and attributes. From 
experts' feedback, we introduced some restrictions to remove questions
that are blatantly nonsensical (e.g., university librarians cannot be
senders of content posted on online learning systems). Following these
restrictions, we ended up with a total of \change[Yan]{2,115}{1411} questions. \remove[Yan]{Nevertheless, since we cannot ask a total of 2,115 questions in one survey session, we decided that we would first, }We randomized  the questions and \remove[Yan]{second, } divided them up into \change[Yan]{24 surveys}{15 sets} (\change[21]{15} 12 with 88 questions, 3 with 89 questions) with about 30 respondents each. That way, we would be able to ask all possible questions within this context (i.e., achieve completeness) at a reasonable cost (\$2 per user per survey, plus AMT fees).

Some of the remaining questions, while perhaps valid, are not applicable in the real world and thus make little sense to the \change[Yan]{survey}{survey} participants  (e.g., it might be unlikely for certain senders to have access to certain attributes). We therefore provided users with three different answer options that suggest nonsense (i.e., ``Does not make sense" (DMS) questions):

 \emph{1) The sender is unlikely to have the information}
 
 \emph{2) The receiver would already have the information}
 
 \emph{3) The question is ambiguous}.

In total, we had 451 respondents to the 15 surveys: each user had to respond to
88-89 questions, with 28-32 respondents per question in each
survey. The average completion time per user and survey was around 14
minutes.

\subsection{\change[Yan]{Evaluation rule extraction}{Approximation of Users' Privacy Expectations}} 

In this section we look at the \textit{pulse} functionality of the framework. We introduce a number of indicators that together allow us to construct an estimate of the users' overall attitude towards the existing set of privacy norms. Specifically, we considered three metrics in
our evaluation: the \emph{norm approval score}, the \emph{user approval score} and the \emph{divergence score}.
\paragraph*{Norm approval score (NA)}
Our most important metric is the \emph{norm approval score (NA)}.
This is our measure of what question is approved by the community for
the operational privacy-rule set.  We define the NA score of question $i$ as follows:

 \begin{gather}
 \mathit{NA}_i= \frac{\sum_{j=1}^{m} Y_{i,j}}{\sum_{j=1}^{m} {(Y_{i,j}+N_{i,j}+\mathit{DMS}_{i,j})}}= \frac{\sum_{j=1}^{m} Y_{i,j}}{m}
 \end{gather} 
Here, $Y_{i,j}$ is defined to be 1 iff respondent $j$ answered ``Yes'' to 
question $i$. Similarly, $N_{i,j}$ and $\mathit{DMS}_{i,j}$ indicate whether user
$j$ answered ``No'', respectively, chose ``Does not make sense''. Thus,
$\mathit{NA}_i$ is the ratio between the total number of ``Yes'' answers and the
number of all answers for question $i$ across all $m$ respondents.
A norm is considered approved if its NA exceeds a certain threshold, e.g., a simple majority ($>50\%$). 
\paragraph*{User approval score (UA)}
This metric measures the relative number of norms that have been
approved by a given respondent. Formally, the value $\mathit{UA}_j$ for respondent $j$ is
defined as
\begin{align}
\mathit{UA}_j = \frac{\sum_{i=1}^{n} Y_{i,j}}{\sum_{i=1}^{n} {(Y_{i,j}+N_{i,j}+\mathit{DMS}_{i,j})}} = \frac{\sum_{i=1}^{n} Y_{i,j}}{n}
\end{align}
where $n$ is the total number of questions in the survey that $j$
responded to. 

\paragraph*{Divergence score (DS)} This metric looks at how the
answers of individual respondents vary from the norms that have been
approved and disapproved  by the whole community subject to a given NA
threshold. Intuitively, it quantifies how dissatisfied a user is with
the extracted set of operational norms. Formally, the divergence score
$\mathit{DS}_j$ of respondent $j$ is defined as
\begin{align}
\mathit{DS}_j = \sum_{i=1}^nc_i \oplus  u_{i,j}
\end{align}
Here, the bit $u_{i,j}$ is defined to be 1 iff respondent $j$ approved
the norm described by question $i$ and $c_i$ is defined to be 1 iff
the community as a whole approved the norm.  Hence, $\mathit{DS}_j$
indicates the number of times respondent $j$'s expectations differed
from the operational privacy rule set that was enforced based on the
chosen NA threshold.

 \begin{figure*}
\centering
\begin{minipage}{0.45\textwidth}
%\hspace{-3em}
\centering
   \includegraphics[scale= 0.45]{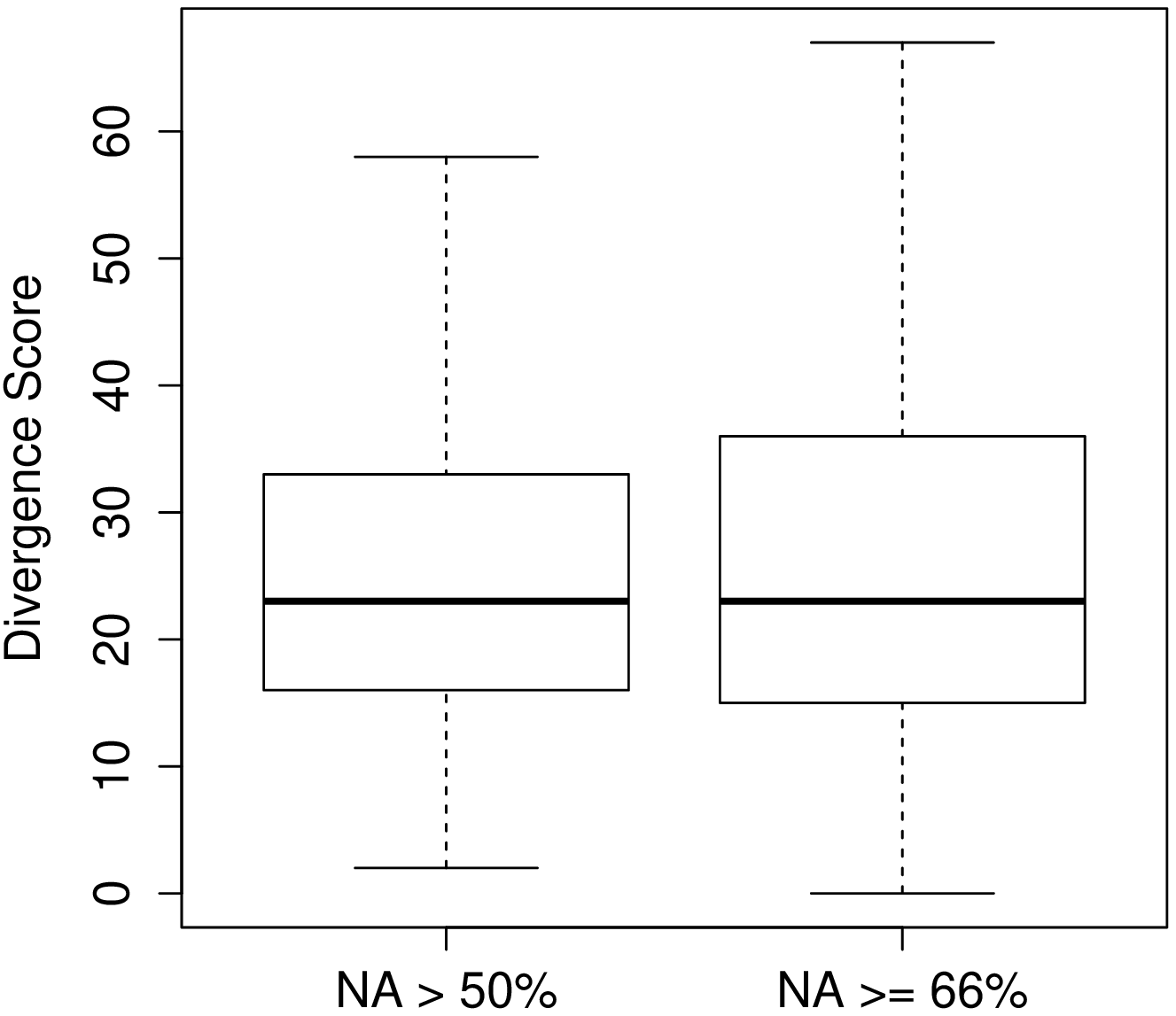}
    \caption{Divergence Score for 50\% and 66\% NA  thresholds: depicts how individual users' DS varies with the respective NA threshold values.}
    \label{fig:BoxplotDivergence5066}\end{minipage}\hfill
\begin{minipage}{0.45\textwidth}
%\hspace{-2em}
\centering
      \includegraphics[scale= 0.4]{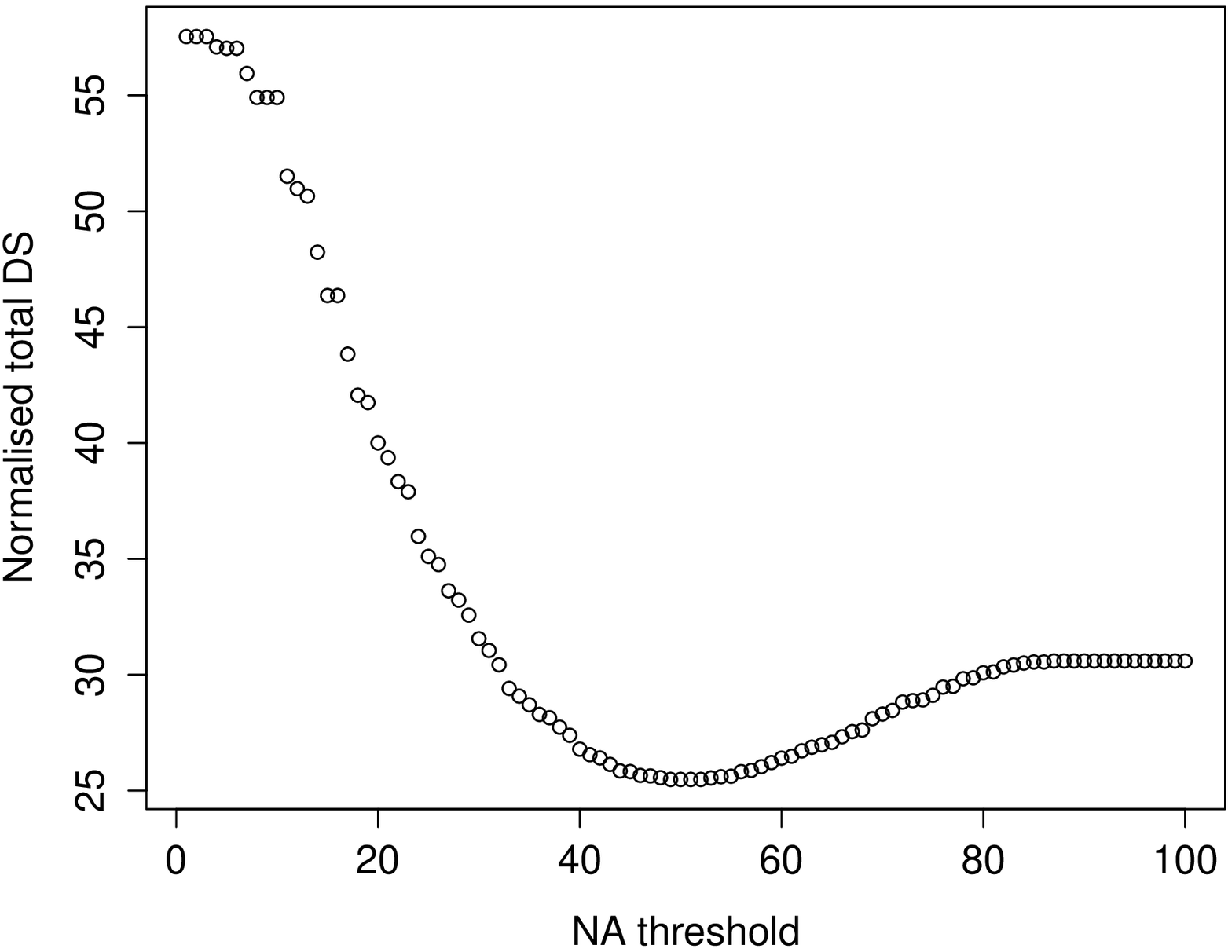}
    \caption{Total DS across all possible thresholds: for each NA threshold we calculated and aggregated the DS score for each of the users. The final number was normalized by the number of the users.}
    \label{fig:DivergentMetric}
\end{minipage}
\end{figure*}

\subsubsection{Summary of crowdsourced data}

We summarize relevant results in Table~\ref{tbl:Approval}. The number behind the ``Yes'' and
``No" columns reflects questions exceeding the respective
NA thresholds. In 36 questions, the respondents could not reach any agreement because the number of
``Yes'' and ``No'' answers was identical. Although this is a small percentage of the total number of questions, it highlights an
important point: some information flows require closer attention;  if this is the case our design allows individuals to select norms according to their personal preferences and identify points of contention through formal verification techniques.

\begin{table}
%\scalebox{0.95}{
%\begin{minipage}{0.5\textwidth}
\centering
\begin{tabular}{cccccc} 
\toprule 
  & \textbf{Yes } &  \textbf{No } & \textbf{Total} \\
\toprule 
\textbf{NA  $>50$}  &  
315  &  645 & 960 (68\%)  \\
\midrule 
\textbf{NA $ \geq 66$}   & 115 & 300 &  415 (29\%)  \\
\toprule
\textbf{Yes = No }&     &     & 36 (2.6\%)\\
\end{tabular} 
\caption{Summary of approved and disapproved norms across the surveys. }
\label{tbl:Approval}
%\end{minipage}}
\end{table}

\subsubsection{Norm approval thresholds}
Next, we analyzed \remove{the capability of the approach to reach a consensus
for} the different NA thresholds and how these threshold choices affect the
users' approval and divergence scores. We focused first on the two
thresholds of 50\% and 66\%.  Figure~\ref{fig:BoxplotDivergence5066}
depicts two boxplots for all users across all \remove{DMS-3 type} questions for
the two NA thresholds. Both populations look very similar. To verify
the difference in means of these two population we ran a one-way ANOVA
to test a \emph{null} hypothesis that there is no difference between
the populations of means under different thresholds. We can reject the
null hypothesis with significance level $p=0.000165$ ($p<0.05$).  A
Tukey HSD test identified that using the 66\% threshold increases the
DS score by 4.8\%. 

In other words, this shows that the 66\% threshold results in a higher disapproval among users with regards to their expressed privacy expectations.

\subsubsection{How satisfied are users with the set of chosen norms?}

Figure~\ref{fig:Divergence} depicts a scatter
plot that shows the number of users with the same DS score across all
\change{DMS-3 surveys}{the questions} for an NA threshold of 66\%.  The plot indicates a large
concentration of respondents with a relatively small DS. This means that,
overall, the users \add[Yan]{in our polls} are satisfied with the operational privacy rule set
chosen by the system for this specific NA threshold.

Furthermore, to understand how the DS varies across all the different thresholds,
we calculated a combined DS for all possible NA thresholds (0\% to 100\%)
and normalized it by the number of total users that had taken the
survey.  The normalization provides us with the combined DS score of
all users per threshold.  Our results, depicted by
Figure~\ref{fig:DivergentMetric}, show that when the threshold is at
its minimum, DS is at the maximum. Recall that DS represents the level
of dissatisfaction of users. We can therefore interpret this result as
follows: when the threshold is low, more questions are approved,
meaning that a significant number of privacy rules that users prefer
to disapprove are included in the operational set. The lowest DS values 
are in the 40\% to 60\% NA threshold range. The best candidates for an
actual threshold choice,   \add[Yan]{for this specific population based on their feedback,} therefore seem to lie in that
range. Interestingly, the DS converges around the 35 mark from 66\% to
100\%. This shows that, in our \change[Yan]{surveys}{polls}, more people opt to disapprove
norms than approve them.

\begin{figure*}
\begin{minipage}{0.45\textwidth}
%\hspace{-2em}
   \includegraphics[scale= 0.4]{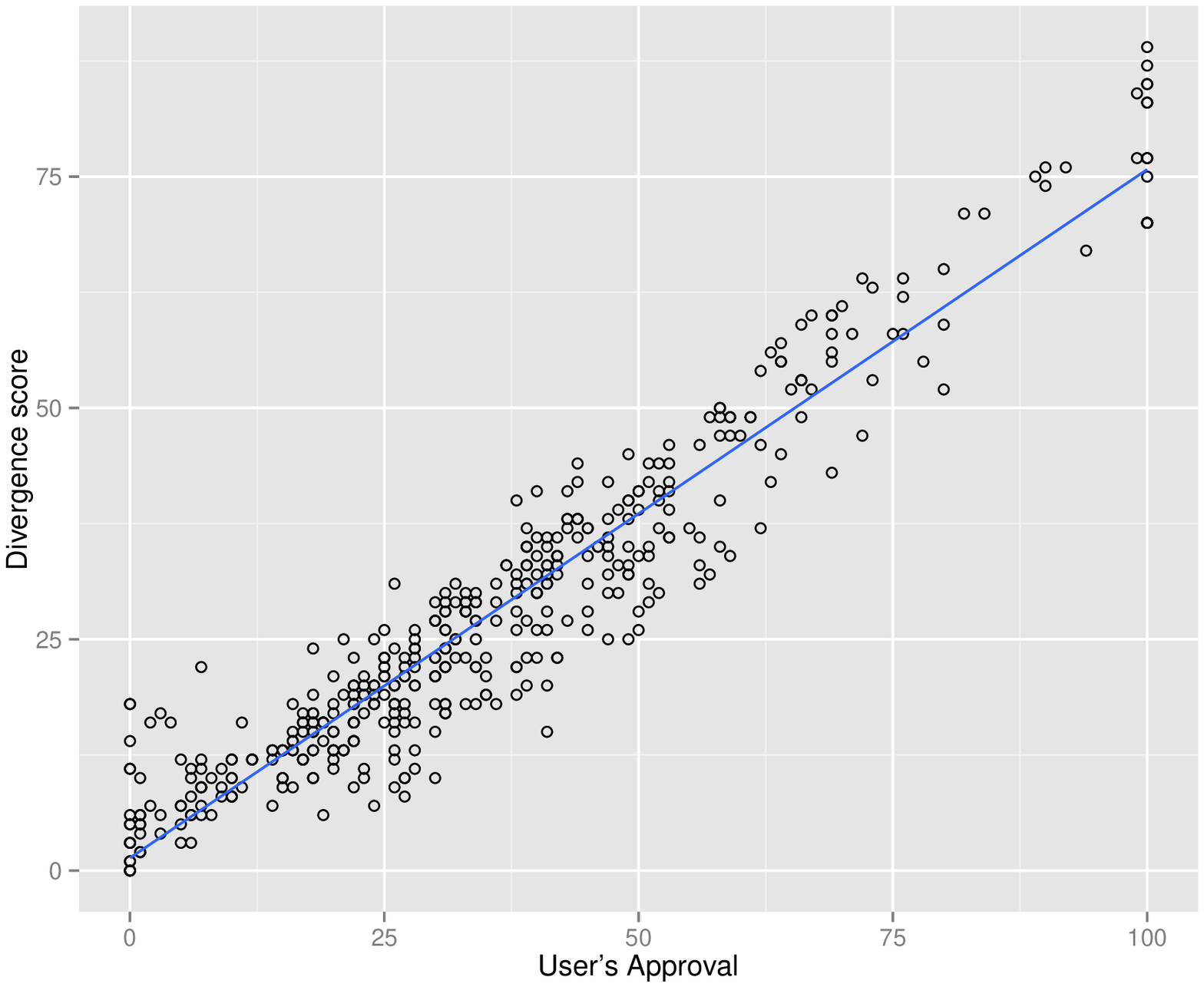}
      \caption{Scatter plot of Users' Approval and Divergence Score
        for each user for NA threshold 66\%.}
       \label{fig:ScatterPlotDSAM66}
\end{minipage}\hfill
\begin{minipage}{0.45\textwidth}
%\hspace{-2em}
\includegraphics[scale= 0.4]{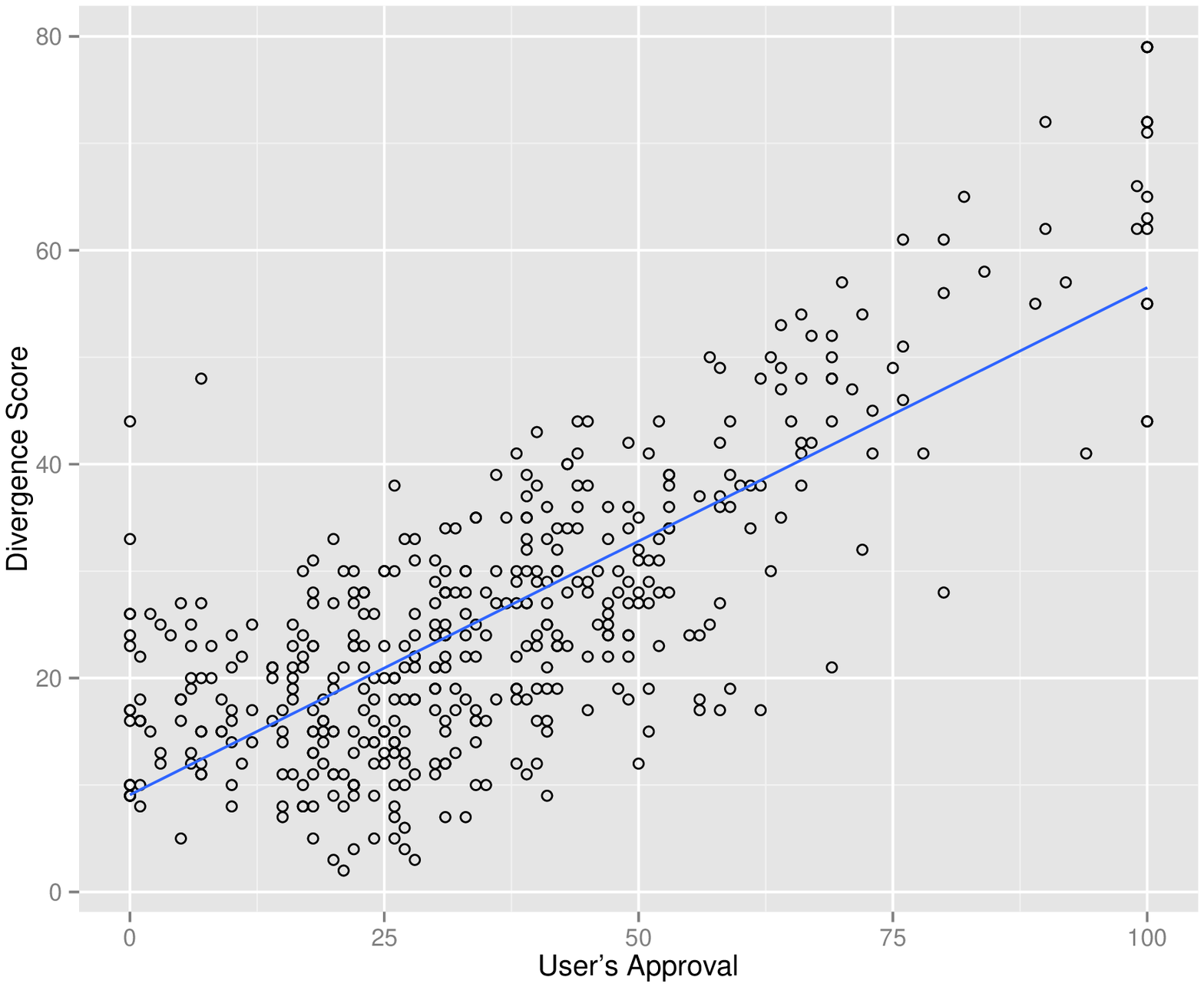}
\caption{Scatter plot of Users' Approval and Divergence Score for each
  user for NA threshold 50\%.}
\label{fig:ScatterPlotDSAM50}
\end{minipage}
\end{figure*}
%\item[AM vs DS.] 
\begin{figure}
      \includegraphics[width= 0.5\textwidth]{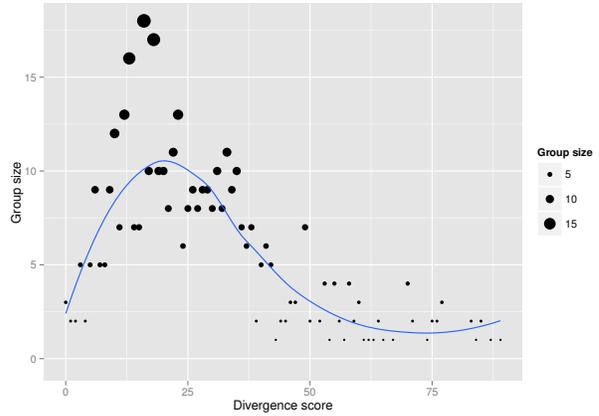}
    \caption{Scatter plot for divergence score}
    \label{fig:Divergence}
\end{figure} 

%\subsubsection{UA vs.\ DS}
\subsubsection{Individual privacy expectations vs Social norms}
Figure~\ref{fig:ScatterPlotDSAM66} shows that there is a linear
relationship between UA and DS for the individual users for a 66\% NA
threshold. Linear regression analysis confirms this ($r^2$ = $0.87 $,
formula: $\mathit{DS}=0.69*\mathit{UA}+2.044$). The 66\% NA threshold makes it hard to
approve privacy rules; users with a very high UA score will often be
disappointed, thus having higher DS. Conversely, users that have a lower
UA are more likely to agree with the community rules.  We can
observe a similar pattern with an NA threshold of 50\% on
Figure~\ref{fig:ScatterPlotDSAM50}; however, relative to the 66\%
threshold, user satisfaction is slightly higher as more privacy rules
are approved on average.

%\end{description}
\subsection{Verification of extracted rules}\label{lbl:verificationTest}

Finally, we evaluate the effectiveness of formal verification
technology to analyze the consistency of the derived privacy logic. We
used the theorem prover Z3 to check whether the \add{crowdsourced}
rules \remove{extracted from the survey data }guarantee certain
privacy properties by encoding both the rules and the properties into
an Effectively Propositional Logic (EPR) as described in
Section~\ref{lbl:verification}. Specifically, our goal was to assess
whether we can use Z3 to automatically check the consistency between
the rules that we derived from the crowd-sourced data for a chosen
threshold, on the one hand, and to check for consistency violations,
on the other hand. We focused our attention on two specific
consistency properties:

\paragraph{1. Semantic consistency  rules. } This property specifies
that the information flow of each disapproved norm is indeed excluded
from the flows that are allowed according to all approved norms. Note
that this property is not trivially satisfied as the approved and
disapproved norms are not necessarily mutually exclusive. In
particular, the roles of a context are not guaranteed to be disjoint,
e.g., an actor in the classroom context may be both a department chair
and a professor. Thus, we may have situations where a specific flow is
approved if the sender is a professor but disapproved if the sender is a
department chair. Such inconsistencies hint at hidden assumptions of
the survey participants that are not adequately reflected by the
formal privacy rules. Our verification approach allows us to detect
such inconsistencies and subsequently eliminate them by refining the
formal rule model and the survey questions appropriately.

\paragraph{2. Consistency of transitive flows.} This property
specifies that the approved norms are transitively closed. For
example, if a professor is allowed to send the grade of a student to
the registrar and, in turn, the registrar is allowed to send the grade
to a graduate school to which the student has applied, then the
professor should be allowed to send the grade directly to the graduate
school.  A violation of the transitivity property hints at a possible
mismatch between the survey participants' privacy expectations and the
logical implications of their individual choices regarding which
privacy norms should be approved. Using our verification approach, we
are able to detect such violations (respectively, prove their absence)
for arbitrarily long sequences of information flows.

In the following, we describe the experiments we conducted to check for each
of these two properties as applied to the set of norms that we derived from our survey
data. Note that in both experiments the entire verification process
including the norm extraction, the logical encoding of the norms and
properties, and their verification, was fully automated.

All the experiments were run on a laptop computer equipped with an
Intel Core I5 CPU at 2.67GHz and 4GB RAM running Ubuntu Linux. The
running time for each of our experiments was less than 5 seconds. The
memory consumption was negligible.

\subsubsection{Detecting semantic inconsistencies of norms} 

For our experiment, to detect semantic norm inconsistencies we chose
the 50\% threshold
to % specific threshold (TODO: describe/justify choice) to
determine which norms are approved according to the crowd-sourced
survey data.
%~\footnote{Similar procedure is repeatable for any other threshold}
For this threshold, as depicted by Table~\ref{tbl:Approval},
\change{587 (total approved norms with DMS-1 and DMS-3)}{315} of our
total \change{2115}{1411} norms were approved. We then encoded these
approved norms into an EPR formula and used Z3 to check for each of
the \change{1528}{1096} remaining disapproved norms whether the
corresponding information flow was indeed prevented by the approved
rules. Each disapproved norm was checked by sending a separate
satisfiability query to Z3.

Intuitively, semantic norm inconsistencies can only arise if an agent
takes on more than one role in a context at the same time. We
confirmed this intuition by conducting an experiment where we verified
the absence of inconsistencies under the assumption that all roles are
pairwise disjoint. Indeed, under this assumption we were able to prove
that 100\% of the disapproved norms were consistent with the rules for
the approved norms.

To detect actual semantic norm inconsistencies, we considered a model
that took the relationships between the different roles in a classroom
context into account. For example, a TA may also be a student and a
department chair is always a professor. With the realistic model, we
detected that 138 of the 1096 disapproved norms were \add{not} ensured
by the approved norms. For example, one of the violated disapproved norms pertained to a professor sharing a student's test result with other students. Such an information flow was permitted by one of the
approved norms, which allowed a professor to share a test result with
a TA. Since a TA may also be a student, the disapproved norm was indeed
violated.

%\subsubsection{Realistic role semantics} 

There are a number of possible ways in which such violations could be
resolved (e.g., by refining the privacy rules or domain ontology).
These are outside the scope of this paper. The focus of our experiment
was to demonstrate that we can automatically detect all such
violations, or alternatively prove their absence.

\subsubsection{Detecting inconsistencies in transitive flows} 

The final experiment was designed to check for inconsistencies due to
transitive flows. Similar to the previous experiment we encoded the
logic into an EPR formula and used Z3 to check for any violations of
the transitivity property. The transitivity property involves
reasoning about arbitrarily long chains of information flows. This
means that for a specific set of approved norms, the number of
concrete chains of information flows that are consistent with the
rules but violate transitivity may be infinite. However, we observed
that for any specific violation, there always exists a \emph{similar}
violation involving a chain of bounded length. This means that all
transitivity violations can be classified by a \emph{finite} set of
small violations. This observation allowed us to exhaustively
enumerate all types of transitivity violations for a given set of
approved rules. To do so, we used Z3's model generation capability to
generate models that witness a small violation of transitivity.

For the 66\% threshold, where 115 of our total 1411 norms were
approved, we automatically detected 59 transitivity violations. On
closer inspection, we found that one such violation was the result of the
following two approved norms:

\begin{enumerate}
%\emph{"TA","department_chair","attendance","2",17
\item  \emph{A TA is allowed to send information about a student's
    attendance to a professor if the student is performing poorly.}
\item  \emph{A professor is allowed to send information about a student's
    attendance to the department chair if the student is performing poorly.}
\end{enumerate}
However, a TA was not allowed to send the attendance information
directly to the department chair, leading to a violation of
transitivity. The approval rate of this rejected norm was only
17\%. Contrasted with the high approval rates of more than 66\% for
the two approved norms involved in the above transitive flow, this
discrepancy hints at a possible violation of the actual privacy
expectations of the users.

\iffalse
 Note, the violation does not
mean that anything is particularly wrong the enforced policies, in
fact, it is consistent with the CI theory; it serves as a notification
of such possibility and leaves it to the enforcers of the policies to
handle these in the manner they see fit, e.g., either by removing or
adding constraints.
\fi

\iffalse
\subsubsection{Summary} 
This shows that formal verification techniques can be used \add{as a \emph{tool}} to
automatically prove privacy properties of the derived norms. \change{On the
other hand, it shows that}{Furthermore,} formal verification can also help to
identify imprecisions in the \remove{survey} questions from which the privacy
rules are derived. These imprecisions can then be weeded out in an
iterative refinement process.

\fi

%!TEX root = main.tex
\section{Discussion}\label{sec:discussion}

{\bf Broader applicability:} We note that we only discussed our framework in a specific case of an educational context. %In principle, the same approach can easily be applied for other context definitions or a much broader/expanded set of actors, attributes or transmission principles.  %We note that what we have presented in our evaluation is only a specific example instantiation of our CI-based privacy framework for a specific case of the educational context. 
In principle, the same approach can easily be applied for other context definitions or a much broader/expanded set of actors, attributes or transmission principles. In addition, the same methodology can be applied in an incremental fashion when new actors or new attributes or new transmission principles are added to the context definition. 

%The core of our proposal of crowdsourcing users' privacy expectation goes against the intuitive way of capturing and defining privacy. Current ways of capturing and enforcing privacy has been embedded in the pre-online revolution. The way we experience privacy today is typically represented by the votes of a selected few. As a society, we delicate this ``chore'' to an authoritarian figure or forced into an obscure agreement without ever knowing whether the derived policies correspond to our personal notion of privacy.  

\noindent\textbf{Choosing real users in practice:}
When applying the CI-based framework in a real-world social platform, we envision that the questions generated by the framework for a given context will be answered by the users of the same platform within the same context. %This will ensure in deriving an appropriate definition of a privacy logic for the specific users within the chosen context. 
For example, consider a set of users who are within a given community in an online social platform. %One can envision using the CI-based framework and using the users within the community to derive a specialized logic for that community. 
In our evaluations, we used AMT to primarily simulate the responses of actual system users. Although previous research~\cite{lin2012expectation,martin2014privacy,ismail2015crowdsourced} provides an early affirmation of the effectiveness of crowdsourcing tools, suggesting that large-scale surveys can indeed be effective for discovering norms, we interpreted our survey only as an approximation of the kinds of feedback we would receive in a mature system with actual users. We would also like to emphasize that, in an operational system, users would only have to respond to a significantly reduced number of questions. Finally, AMT does not allow us to test cases in which new norms are introduced, as described in Section~\ref{sec:evolution}. We aim to address this issue in future work.

\noindent\textbf{Privacy logic.} We acknowledge that technical systems usually embody an idea of what privacy is, whether this is merely implicit in the design or stated explicitly in theoretical terms, as we have done. The sources of such ideas may be varied: the intuitions of a design team, law and other regulatory systems, privacy experts, etc. We do not take issue with these sources. However, the approach we take, we believe, is particularly well suited for information systems involving diverse social actors interacting with one another through complex patterns of communication (or flows) of information.

\noindent\textbf{Verification.} The encoding into EPR and subsequent verification is fully automated. Our approach differs from prior work in that our encoding of norms and properties remains within a decidable logic that admits practical decision procedures. In particular, this means that a failed verification attempt is always due to an actual property violation (as opposed to an incompleteness in the verification approach). What is more, if verification fails, the theorem prover creates a model from which a violating information flow can be extracted. This enables the automated diagnosis of inconsistent norms, which can in turn be used to automate a feedback loop in the crowdsourced norm generation.
\section{Related Work}\label{sec:related_work}
In this section we acknowledge important prior and adjacent work that is related to ours.  This roughly falls into four different categories: a) Other Access Control (AC)  frameworks; b) Other efforts adopting the CI framework for building privacy-preserving systems; c) Other compliance-based approaches to privacy-by-design (PbD); d) Other efforts utilizing crowd-sourcing to promote system privacy.

%\subsection

\paragraph{Other AC frameworks.}

Many other frameworks have been designed to manage access control to information and other privacy-preserving functions. 
Barth et.al.,~\cite{barth2006privacy}, provide a comprehensive comparison of the CI  framework to other existing models such as Role-Based Access Control (RBAC), the eXtensible Access Control Markup Language (XACML), Enterprise Privacy Authentication Language (EPAL), and the Platform for Privacy Preferences (P3P). 

\begin{description}
%\subsubsection{RBAC}
\item[RBAC.] In RBAC, access control is defined in terms of available resources and users' roles that have access to it. Compared to RBAC, CI is a more generic framework that introduces additional attributes, namely contexts, subjects, as well as transmission principles, that capture more accurately the different dynamics according to which information is shared. 
%\subsubsection{the eXtensible Access Control Markup Language (XACML) } 
\item[XACML.]
The eXtensible Access Control Markup Language (XACML) is a generic XML-based markup language that allows  specifying attribute-based AC policies for resources.  XACML can be used to implement Attributed  Base Access Control (ABAC) as well as the RBAC scheme. We can also express the CI framework using XACML  by capturing the  contextual and information flow semantics behind the CI norms.
%\subsubsection{Platform for Privacy Preferences (P3P)}  
\item[P3P.] The P3P language is designed to help websites describe their privacy polices with regards to users' data. In contrast to the CI framework, P3P is limited to policies involving only two parties, i.e., the website and the visitor, operating in a very specific, global context~\cite{barth2006privacy}.  
\end{description}

%\subsection

\paragraph{Other works based on CI.}

In \cite{barth2006privacy} the authors proposed a logical framework for reasoning about
privacy expectations and privacy practices using CI. The same  formalization of contextual has been also used in \cite{DBLP:conf/trustbus/LamMS09}. The framework uses first-order temporal logic (FOTL) to express norms by describing  the actors that participate in each context, the roles these actors play, and their knowledge states, all at a specific point in time. While FOTL-based formalisms can model temporal properties related to contextual integrity, the logic itself is too expressive to serve as a suitable foundation for tools that mechanize reasoning about
privacy norms in real-world systems. For example, the valid formulas of full FOTL are not even recursively enumerable. Consequently, any automated approach to reasoning about validity, respectively, satisfiability  of formulas in full FOTL is inherently incomplete~\cite{Abadi87}.

In \cite{criado2015implicit} the authors propose computational and information
models of Implicit Contextual Integrity in a social network.
The main motivation behind this work is based on the idea that
context in OSN is never explicitly defined and must be inferred
from the information itself. The paper introduces the notion of
an Assistant Agent which uses the defined information model
to infer any implicit contexts, relates them to existing norms
and stops undesirable information flows. In our work, we rely on the users to decide on valid and relevant information flows and contexts in the system. 

Similarly, Y. Krupa and L. Vercouter~\cite{krupa2012handling}  looked into having an Assistant Agent be part of a
Privacy as Contextual Integrity for the Agent Systems (PrivaCIAS)
framework for open and decentralized virtual communities.
The agent is designed to assist users with preserving their
information privacy as well as detecting other users that violate
the established privacy norms. The framework is based on a trust
model where agents gossip about their ``experiences'' to inform
other agents of any violations as well as provide passive
feedback. The ultimate goal is to socially exclude ``bad'' agents
from the system. While this work focuses on enforcing  CI
norms in a distributed environment, it does not discuss how to
extract the prevailing set of norms in the first place. Our  framework can complement such initiatives by providing a mechanism for  extracting CI privacy norms from the community of users and verify their consistency. 

%\subsection
\paragraph{Compliance-based approaches to PbD.}

Rising to the call for PbD, numerous approaches have been invented to map system constraints to privacy policies that are generated external to these systems. Two examples, are Sen, et. al. in \cite{sen2014bootstrapping} and Breaux and Anton in~\cite{breaux2008analyzing}. Sen, et. al. have developed LEGALEASE, a language with precise semantics that enables enterprises to express privacy policies against which practices can be checked for compliance. With similar goals, Breaux and Anton have proposed a methodology for extracting implementable software requirements from privacy and security regulatory requirements expressed in legal language that is sometimes vague and ambiguous. The intention in both examples is similar to our own in that it seeks to translate privacy requirements expressed in natural language (assuming this covers legal language) into formally expressed rules that allow for compliance checking. The threat scenarios, however, are quite different. Further, both examples start with externally generated privacy norms, which they seek to express in formal language.
% \vspace{-10pt}
%\subsection
\paragraph{Other uses of crowdsourcing to generate privacy rules.}
One effort that is quite close in spirit to our own
is~\cite{sadeh2014modeling} which seeks to ease the burden on users
when tailoring privacy policies on mobile apps to accurately reflect
their privacy preferences. The system clusters users according to
their willingness to share information with app providers and
configures settings on future apps based on the position in a cluster. Relevant differences are (i) that it applies to a dyadic relationship between the user and app provider, and (ii) it seeks to model preferences while our work aims to model social norms.  

Similarly Tohn~\cite{toch2014crowdsourcing} has proposed the SuperEgo
system, which uses crowdsourcing to enhance location privacy
management in mobile applications. SuperEgo uses the perception of the crowd to predict the privacy preferences of an individual. The system relies on a crowd-opinion model and a mixture of decision-making strategies to classify  the information as private or not. Although this work is conceptually similar in that it uses crowdsourcing to infer relevant privacy policies for the user, it is limited to a location-based privacy context. As noted by the author, the CI framework is more expressive and capable of capturing privacy-rules in a range of different contexts. 
\medskip\\
In summary, previous contributions focused on applying the CI framework to
enforce privacy norms in different domains. Our work builds on these
efforts by capturing relevant norms in a given context and then
proposes a privacy-preserving framework that efficiently encodes them
into a actionable and verifiable privacy logic. %Finally, our approach allows us to verify the selected privacy rules for consistency.
%!TEX root = main.tex
\section{Summary and Conclusion}\label{sec:conclusion}

In this paper we described a framework  for discovering verifiable and actionable privacy norms in a community of users based on the theory of contextual integrity.  

We evaluated our proposed framework by conducting an extensive survey involving more than 450 participants and 1400 questions to derive a set of privacy norms in the educational context. We were able to show that the Datalog encoding of the derived norms enables us to automatically verify the consistency of (transitive) information flows and automatically detect logical inconsistencies between individual users' privacy expectations, on the one hand, and the derived privacy logic, on the other hand. 
Our results leave us optimistic about the feasibility of a full-fledged information system that operates based on the design principles of crowdsourcing, formal verification, and contextual integrity. 

Future work includes an in-depth investigation into more elaborate approval and divergence metrics, an extension of our design to handle inter-domain privacy rules as well as the release of a prototype system based on privacy norms discovered using the methods we have developed.

Looking even further into the future, our work paves the way towards information systems that operate on a foundation of substantive privacy rules that reflect the rough consensus of given communities. These could include communities across the domains of education, health, or more general social domains. The mechanisms we have developed for extracting, expressing, and validating a set of common rules could be integrated into such systems. By incorporating these mechanisms into information/social systems, user feedback can be continuously elicited, which will enable a system to refresh rules continuously to reflect evolving community norms and standards.

{\footnotesize \bibliographystyle{acm}
\bibliography{main}}

\end{document}